# Sentiment Analysis on IMDB Movie Comments and Twitter Data by Machine Learning and Vector Space Techniques

İ.Tarımer[1], A. Çoban[1*], A.E. Kocaman[1]

[1]Department of Information Systems Engineering, Technology Faculty, Muğla Sıtkı Koçman University Muğla, Turkey

*Abstract*

*This study's goal is to create a model of sentiment analysis on a 2000 rows IMDB movie comments and 3200 Twitter data by using machine learning and vector space techniques; positive or negative preliminary information about the text is to provide. In the study, a vector space was created in the KNIME Analytics platform, and a classification study was performed on this vector space by Decision Trees, Naïve Bayes and Support Vector Machines classification algorithms. The conclusions obtained were compared in terms of each algorithms. The classification results for IMDB movie comments are obtained as 94,00%, 73,20%, and 85,50% by Decision Tree, Naive Bayes and SVM algorithms. The classification results for Twitter data set are presented as 82,76%, 75,44% and 72,50% by Decision Tree, Naive Bayes SVM algorithms as well. It is seen that the best classification results presented in both data sets are which calculated by SVM algorithm.*

## 1. INTRODUCTION

Firms with influence of rapidly developing technologies have taken social media to their targets because they are inadequate to reach the customers with ordinary marketing methods. For this reason, social media has taken part as the most important one among the data sources. Data taken from social media take part play a decisive role for rivalry at trade. Because the comments of customers about a product or service can be viewed everywhere via social media [1]. Almost every day, a lot of positive and negative comments are shared about the issues of products, services, brands, institutions, economics, politics, sports and so on in social media. These opinions and comments which are spread in the social media on the internet, show how successful the products and services are, as well as direct marketing rivalry and the economy. The influence of these comments to the economy has led to the need for institutions to follow trends in social media. As a natural consequence of this, it has led several studies of emotion analysis in the social media.

The shared contents at social media (Facebook, Twitter, Linkedin and Google+, etc.) are in kinds of text, pictures or videos. However, there are no tag meanings of the shared contents such as topic, positivity, negativity. For this reason, the perception is that the topic is stored in the mass of data shared in social media. [2]. When the big unprocessed pure data is to be examined one by one, it is very harsh to understand them with human perception. Therefore, processing of data by natural language processing methods has emerged as a solution [3].

When social media data is analyzed within its unprocessed form, it appears that the vast majority of them are seem that they are composed from erroneous words, abbreviations, and social media specific jargon words that are not used in everyday speech; therefore, it is quite harsh working on these [4]. Nevertheless, many researches have been done on the data taken from Twitter by natural language processing, machine learning and data mining techniques, and projects for industrial scale have been carried out. [4][5][15].

Nikfarjam and Azadeh have observed that sentiment analysis by SVM on twitter data and forums relevant to health were resulted with 82.1% success in their study [13]. Kaynar, et al., have conducted sentiment analysis study on the data including the film comments in IMDB and obtained the highest values of success by artificial neural networks and support vector machines [14]. Moraes, et al. presented an empirical comparison between SVM and ANN regarding document-level sentiment analysis [16]. In [17], a sentiment analysis was performed by collecting data from the Twitter; their success of classification was about 90%.

## 2. MATERIAL AND METHODS

In this section, the data sets to be worked on and their fields from which are drawn are specified; the methods and techniques of data mining on these data are explained. In the study, 2000 rows about IMDB movie comments and 3200 Twitter data about 4 technology companies were used. The data drawn from the Hashtags present the tweets published in English language about the firms of Lenovo, Samsung, Sony and Apple. To withdraw data from Twitter, RapidMiner software is used. This program is a data annotation software platform developed by the same company that provides an integrated environment for data preparation, machine learning, deep learning, text mining and analytics based on estimation [6]. On the other hand, to do data mining, an open source application which is called KNIME has been used in this study. The software 'KNIME' began to develop in 2004 as a registered product at Konstanz University [7].

### *2.1. Natural Language Processing by Vector Space Method*

A vector space consists of vectors, scalars, and operations of vector addition and scalar multiplication. This produces a new vector [19]. Set of text data can be specified in many different shapes by vector space nodel. The most popular way is to count the term frequency. As alternative ways to this is text labeling (POS tagging), n-gram counting or latent semantic analysis. Alternative ways to the term frequency in text labeling are text labeling (POS tagging), n-gram counting or latent semantic analysis [8].

Advantageous of using of vector spaces are listed as: Data structures that can be processed using linear algebra are obtained; all functions defined for the vectors can also be defined for the texts (i.e. cosine similarity); sorting functions (ranking) can also be executed on texts and it can be worked on a piece of text instead of the whole text [9]. Beside these, it is also needed to deal with the big data containing many features after extraction of feature vectors [9].

### *2.2. Term Frequency and Inverse Document Frequency (TF – IDF)*

Term Frequency is a method that it is used to calculate term weights in a document. TF IDF is a statistic that it aims to reflect its importance to a document within a collection or in a document [10]. Term Frequency is the frequency of the term within the document. It indicates that how many times a word is used within a document [11]. The Inverse Document Frequency attempts to find out number of usage of the document in multiple documents. It also tries to understand whether this is a term or a link (stop words) [18]. There are two critical numbers at TF-IDF calculation. The first is the number of term within the document handled at that moment, the other is the total number of documents containing the term presence in the corpus.

$$f_{t, d} = \text{This is the number of term in the document } \text{tf}(t, d)$$

It can be simply defined as the number of the word in the document. However, it is also possible to define in different forms according to the need for sensitivity.

For logarithmic scaled frequency, it can be written as:

$$1 + \log(f_{t,d})$$

It is seen that many words can be used many times within a big document. The number of a word in a document is calculated by dividing it by the total number of words [12].

$$tf(t, d) = 0{,}5 + 0{,}5 \cdot \frac{f_{t,d}}{max\{f'_{t,d} : t' \in d\}}$$

In this experimental study, feature(s) was extracted by vector space method for sentiment classification. Positive or negative states of the texts with the longest and most common words in the data set were entered manually and the frequency of each term was added to the sentiment condition.

## 2.3. Work Flow Model Steps

The process steps and the workflow diagram that are followed in the work are shown in Fig 1. The data read from the file is passed to the text extraction stage and the vector space creation stage. After that, the classification process is performed and the success of the classification algorithms used is compared.

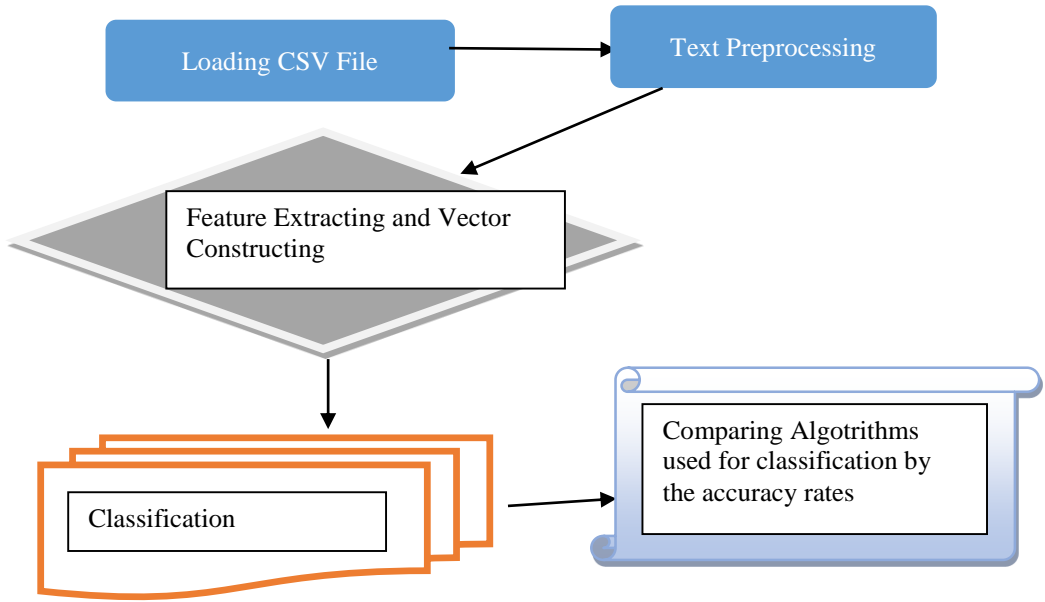

Fig. 1. Work Flow Diagram.

Based on these steps and methodologies, the model shown in Figure 2 is built in Knime.

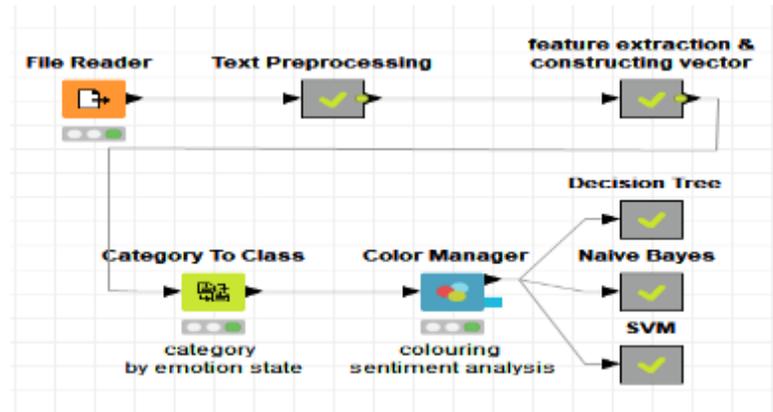

Fig. 2. Constructed Model

The IMDB Movie Comments data has been downloaded from the official website of Knime. Twitter data is downloaded using the Knime interface and saved in the csv file. In this workflow model, 50 twitter texts collected under a hashtag were drawn and recorded. To add generated Csv files to our Knime workflow model with the "File Reader" node in the program, the csv file is read and prepared for processing.

## 2.4. Text Pre-Processing

In this step the prepared and uploaded data files are filtered with some rules. First, by adding the "strings to document" node, the data file is transformed into a process-ready document. Right click on the node and click on "configure" to open the settings menu. For the sections "Title column" and "full text" in "document source column", the text column in which the tweets are located is selected. For "Document category column", the column to be categorized is selected. The negative and positive values specified in this column are removed with the "Punctuation eraser" node; the numeric and stall words are filtered and all terms are converted to lower case. Then, the word roots are removed using the node "Snowball Stemmer". The concept of a real word is found at the beginning and the same information is conveyed to the context of a document classification or subject recognition. In addition to English texts, the snowball splitter node can also be applied to texts in various languages (German, French, Italian, Spanish, etc.). The node uses the Snowball rooting library. The text pre-processing meta node part is given in Fig 3.

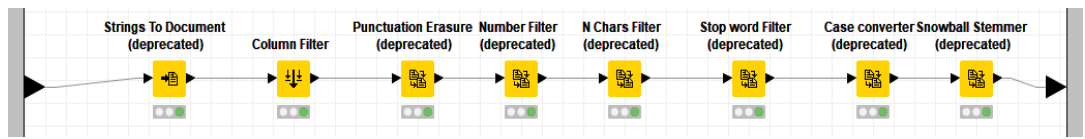

Fig.3. Text pre-processing steps

## 2.5. Feature Extraction and Vector Creation

After this preprocessing to prepare the text, the main focus of this analysis is on the feature extraction and vector generation part. By using the features in the document vectors, the desired terms are extracted and after that the business process is started to classify the documents. Using the "BoW Creator" knot, the words are created one by one by bag. These term words created with the "Term to string" node are translated into strings for the purpose of determining and processing frequency values. These converted terms were passed to the "row filter" node to filter sentiment conditions into rows. The sentiment conditions from the "Row filter" and the term frequency values from the "TF" node were combined with the "refine row filter" node to perform positive unfavorable filtering. The generated filter is transferred to the "document vector" node and the attribute and vector document are created as shown in Fig 4.

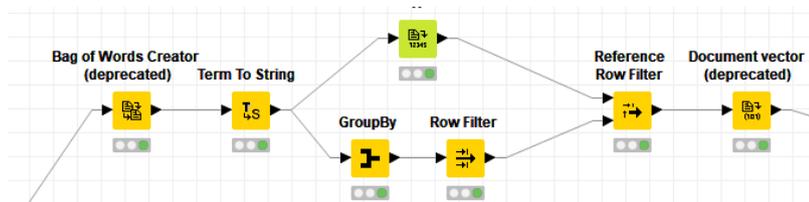

Fig. 4. Feature Extraction and Vector Creation

### 2.5.1. Classification

The emotion classification document results obtained with the "color to manager" node to colorize the "Category to class" node and the negative ones to red and the positive ones to green are given in Figure 5.

| Row ID | S Docum… | Docum… | D galaxi | D phone | D appl |
|---|---|---|---|---|---|
| Row225 | POS | "???Flight … | 0 | 0 | 0 |
| Row226 | POS | "?MASAKI … | 0 | 0 | 0 |
| Row227 | NEG | "@AssnsGl… | 1 | 0 | 0 |
| Row228 | NEG | "@Barbaro… | 0 | 0 | 0 |
| Row229 | NEG | "@BestBuy… | 0 | 0 | 0 |

Fig. 5. Colouring Sentiment Analysis

## 3. MACHINE LEARNING'S SUCCESS SCORE AND STATISTICS

This sentiment classification was concluded with the "Roc curve", accuracy statistics and Confusion Matrix values of emotion classification model by adding Decision Tree, Naïve Bayes and SVM algorithms to the model flow chart in order to display the success score of our model. From the results obtained on the IMDB data, the Roc curve for Decision Tree Learner is shown in Fig 6.

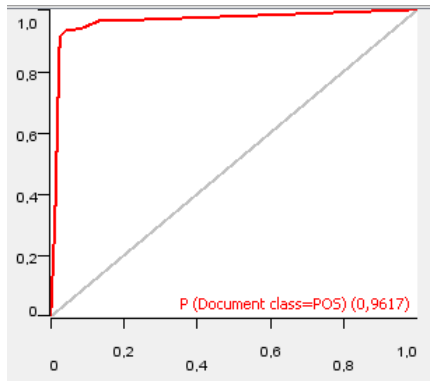 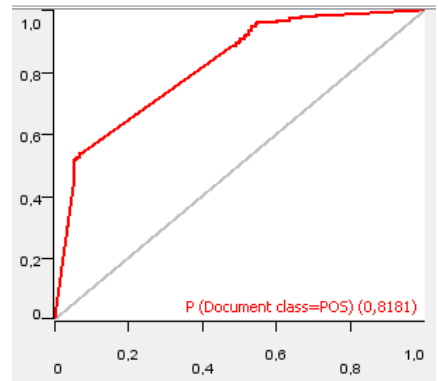

Fig. 6. Decision Tree ROC Curve for class "POS"     Fig. 7. Naïve Bayes ROC Curve for class "POS"

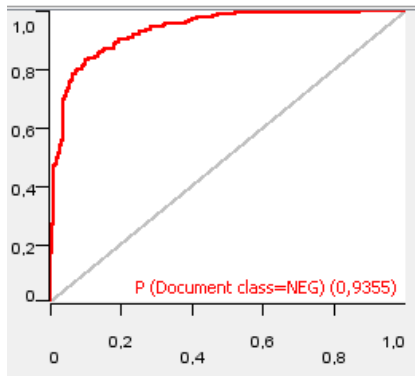

Fig. 8.SVM Learner ROC Curve for class "NEG

For Decision Tree Learner, the area under the ROC curve is calculated as 0,9617. The detailed accuracy rate is shown in Table I.

**Table I.** Decision Tree Detailed Accuracy Rate

|     | TrueP | FalseP | TrueN | FalseN | Recall | Precision | Sensitivity | Specifity | F-Measure |
| --- | --- | --- | --- | --- | --- | --- | --- | --- | --- |
| NEG | 291 | 27 | 273 | 9 | 0,97 | 0,92 | 0,97 | 0,91 | 0,94 |
| POS | 273 | 9 | 291 | 27 | 0,91 | 0,97 | 0,91 | 0,97 | 0,94 |

When we examine the Complexity Matrix results for Decision Tree Learner, 291 of 300 negative negatives were found to be correct and 273 of 300 positive negatives were found to be correct. The accuracy rate obtained was 94%. The Roc Curve for the classification made by the Naïve Bayes Learner algorithm on the IMDB film interpretation data is given in Fig 7. The area under the Naïve Bayes learner Roc curve was calculated to be 0,818. detailed accuracy ratio table is given in Table II.

Table II. Naïve Bayes Detailed Accuracy Rate

|     | TrueP | FalseP | TrueN | FalseN | Recall | Precision | Sensitivity | Specifity | F-Measure |
| --- | --- | --- | --- | --- | --- | --- | --- | --- | --- |
| NEG | 278 | 139 | 161 | 22 | 0,93 | 0,86 | 0,93 | 0,54 | 0,78 |
| POS | 161 | 22 | 278 | 139 | 0,54 | 0,88 | 0,54 | 0,93 | 0,67 |

According to the results of Table 4, 161 of positive judgments and 278 negative judgments were correctly found. The accuracy achieved was 73,2%. The Roc Curve for the SVM Learner algorithm is given in Fig 8. In Figure 8, the area under the Roc Curve for the SVM Learner algorithm is calculated as 0,94. Table V shows the detailed accuracy ratios for the SVM Learner algorithm.

Table III. SVM Learner Detailed Accuracy Rate

|     | TrueP | FalseP | TrueN | FalseN | Recall | Precision | Sensitivity | Specifity | F-Measure |
|-----|-------|--------|-------|--------|--------|-----------|-------------|-----------|-----------|
| NEG | 261   | 48     | 252   | 39     | 0,87   | 0,85      | 0,87        | 0,84      | 0,86      |
| POS | 252   | 39     | 261   | 48     | 0,84   | 0,87      | 0,84        | 0,87      | 0,85      |

According to Table III, 261 of positives and 252 negative of judgements are found correctly by SVM algorithm. The accuracy achieved was 85,5%.

In the next part of the study, the classification results on the data obtained via Twitter are shared. The ROC Curve as a result of Decision Tree for the POS class is shown in Fig 9.

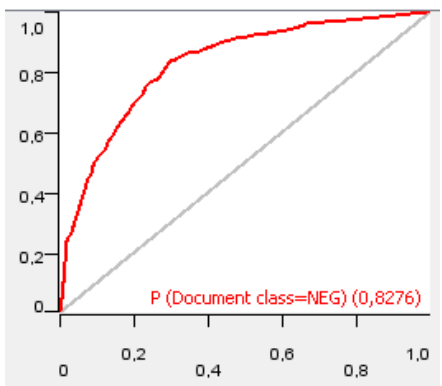

Fig. 1. Decision Tree Learner ROC Curve for class "NEG"

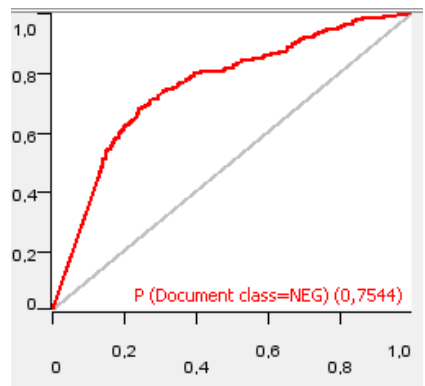

Fig. 10. Naïve Bayes ROC Curve for Class "NEG"

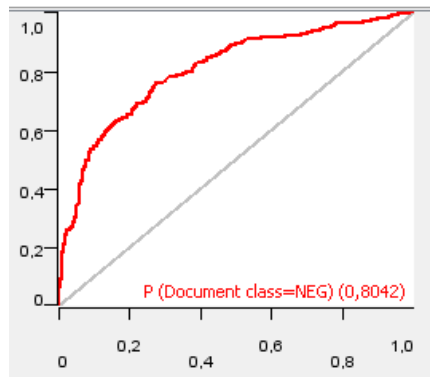

Fig. 2. SVM ROC Curve for class "NEG"

In Figure 10, the area under the ROC curve of the Decision Tree algorithm for the NEG class in the Twitter Lenovo data is found to be 0,8276. Table IV show the detailed accuracy rates for the Decision Tree Learner algorithm.

Table IV. Decision Tree Learner Detailed Accuracy Rate

|     | TrueP | FalseP | TrueN | FalseN | Recall | Precision | Sensitivity | Specifity | F-Measure |
|-----|-------|--------|-------|--------|--------|-----------|-------------|-----------|-----------|

| | | | | | | | | |
|---|---|---|---|---|---|---|---|---|
| NEG | 394 | 125 | 344 | 113 | 0,78 | 0,76 | 0,78 | 0,73 | 0,77 |
| POS | 344 | 113 | 394 | 125 | 0,73 | 0,75 | 0,73 | 0,78 | 0,74 |

According to the results of IV, 344 of the positive judgments and 394 of the negative judgments were found correct. The accuracy rate obtained is 75.6%. Figure 10 shows the ROC Curve of the "POS" class obtained by Naïve Bayes algorithm.

In the classification made by Naïve Bayes, the area under the ROC curve for the NEG class was found to be 0,7544. Table V shows the detailed accuracy ratios for the Naïve Bayes Learner algorithm.

Table V. Naïve Bayes Detailed Accuracy Rate

| | TrueP | FalseP | TrueN | FalseN | Recall | Precision | Sensitivity | Specifity | F-Measure |
|---|---|---|---|---|---|---|---|---|---|
| NEG | 357 | 129 | 340 | 150 | 0,70 | 0,74 | 0,70 | 0,73 | 0,72 |
| POS | 340 | 150 | 357 | 129 | 0,73 | 0,69 | 0,73 | 0,70 | 0,71 |

According to the results of Table V, 340 of the positive judgments were found to be correct, and 357 of the negative judgments were found to be correct. The accuracy rate obtained is 71.4%.

Figure 12 shows the ROC Curve of the "NEG" class obtained by the SVM algorithm.

In Figure 12, the area under the ROC curve is found to be 0,8042 for the SVM Algorithm. Table VI shows the detailed accuracy rates for SVM Learner algorithm.

Table VI. SVM Detailed Accuracy Rate

| | TrueP | FalseP | TrueN | FalseN | Recall | Precision | Sensitivity | Specifity | F-Measure |
|---|---|---|---|---|---|---|---|---|---|
| NEG | 351 | 112 | 357 | 156 | 0,692 | 0,758 | 0,692 | 0,761 | 0,724 |
| POS | 357 | 156 | 351 | 112 | 0,761 | 0,696 | 0,761 | 0,692 | 0,727 |

According to the results of Table VI, 357 of the positive judgments and 351 of the negative judgements were correctly found. The accuracy rate obtained is 72.5%. The accuracy rates of all classification algorithms are shown in Table VII.

Table VII. Classifiers all together

| | Decision Tree | Naïve Bayes | SVM |
|---|---|---|---|
| IMDB Data | 94,00 | 73,20 | 85,50 |
| Twitter Data | 82,76 | 75,44 | 72,50 |

As can be seen from Table XIII, Decision Tree classification algorithm can produce better results than other algorithms. Naïve Bayes produced the lowest accuracy rate for IMDB data set. The Decision Tree algorithm does not achieve the same success in Twitter data as it produces good result in IMDB data.

## 4. CONCLUSIONS AND SUGGESTIONS

As a result of this study, the highest classification success rate for Twitter and IMDB data in the direction of our findings was reached with decision tree algorithm for both data sets.

Due to the low success rate of Twitter data, IMDB data set was lower than IMDB data due to regular and regular writing by movie critics. Twitter's tendency to decline in success rate due to misspellings made in the data. It is obvious that social media data is difficult to work on, and it is obvious that many of them are composed of misleading words, abbreviations and social media specific jargon words which are not used in everyday speech.

When this large amount of data is tried to be examined individually with its unprocessed pure structure, it is very difficult to perceive even with human perception. For this reason, the workflow model was designed and implemented with the aim of scoring the data with natural language processing methods and emotional classification with the reduced vector space creation techniques and machine learning algorithms.

For the model, some changes and updates can be made for higher accuracy and success. For example, a dictionary tagger node can be added in Knime to create a dictionary tag node. Thanks to this label, some words can be held in separate groups and positive words and negative words can be searched in sentences so that model estimation can achieve more successful results.

Another feature that can be added to our model is the filtering of hashtag and tags, which can make more accurate estimation of our workflow model. There is no knot in Knime for filtering tags, but since it is an open source platform, improvements can be made in Knime in java.

**REFERENCES**

[1] İşlek, Mahmut Sami. *The Effects of Social Media on Consumer Behaviors: A Research on Social Media Users in Turkey. MS thesis*. Karamanoğlu Mehmetbey University Social Science Institute, 2012.
[2] Baykara, M. and Gürtürk, U. (2017). Text Sentiment Analysis.
[3] Chen, Hsinchun, Roger HL Chiang, and Veda C. Storey. "Business intelligence and analytics: from big data to big impact." *MIS quarterly* (2012): 1165-1188.
[4] Nizam, Hatice, and Saliha Sıla Akın. "Comparing The Performance Of Balanced And Unbalanced Data Sets In Emotion Analysis With Machine Learning In Social Media." *XIX. Internet Congress in Turkey* (2014). http://inet-tr.org.tr/inetconf19/bildiri/10.pdf
[5] Yeşilyurt, A., and Şeker, S. E. (2017). "Twitter Sentiment Analysis using Text Mining Methods." *YBS Encyclopedia*. http://ybsansiklopedi.com/.
[6] Hofmann, M., & Klinkenberg, R. (2013). "RapidMiner: Data Mining Use Cases and Business Analytics Applications (Chapman & Hall/CRC Data Mining and Knowledge Discovery Series).
[7] Berthold, Michael R., et al. "KNIME-the Konstanz information miner: version 2.0 and beyond." *AcM SIGKDD explorations Newsletter* 11.1 (2009): 26-31.
[8] Erk, Katrin, and Sebastian Padó. "A structured vector space model for word meaning in context." *Proceedings of the Conference on Empirical Methods in Natural Language Processing*. Association for Computational Linguistics, 2008.
[9] Erk, Katrin. "Vector space models of word meaning and phrase meaning: A survey." *Language and Linguistics Compass* 6.10 (2012): 635-653.
[10] Turney, Peter D., and Patrick Pantel. "From frequency to meaning: Vector space models of semantics". *Journal of artificial intelligence research* 37 (2010): 141-188.
[11] Salton, Gerard, Anita Wong, and Chung-Shu Yang. "A vector space model for automatic indexing". *Communications of the ACM* 18.11 (1975): 613-620,
[12] Rubin, Donald B. "Bayesianly justifiable and relevant frequency calculations for the applied statistician." *The Annals of Statistics* 12.4 (1984): 1151-1172.
[13] Nikfarjam, Azadeh, et al. "Pharmacovigilance from social media: mining adverse drug reaction mentions using sequence labeling with word embedding cluster features." *Journal of the American Medical Informatics Association* 22.3 (2015): 671-681.
[14] Kaynar, Oğuz, et al. "Sentiment Analysis with Machine Learning Techniques." *International Artificial Intelligence and Data Processing Symposium (IDAP'16)*, September. 2016.
[15] Can, Ümit, Alataş, Bilal. "Review of Sentiment Analysis and Opinion Mining Algorithms". *Int. J. Pure Appl. Sci.(IJPAS), Volume 3, Issue 1, Pages: 75-111, 2017*.
[16] Moraes, R., Valiati J.F., Gaviao Neto, W.P. "Document-level sentiment classification: an empirical comparison between SVM and ANN". *Expert Syst Appl. Volume 40, Pages: 621–33, 2013*.
[17] Meriç, Meral, Diri, Banu. "Sentiment Analysis on Twitter". *IEEE 22nd Signal Processing and Communications Applications Conference (SIU 2014), ISBN: 978-1-4799-4874-1/14, Pages: 690 – 693*.
[18] Robertson, Stephen, (2004). "Understanding inverse document frequency: on theoretical arguments for IDF". *Journal of Documentation, Vol. 60 Issue: 5, pp.503-520, https://doi.org/10,1108/00220410410560582*
[19] Çiftçi, Süleyman, (September 2015). "Linear Algebra", *Dora Publishing, 1. Edition, Pages 421*.